\documentstyle[sprocl]{article}

\def\gev{\mathrel{\rm GeV}}
\def\tev{\mathrel{\rm TeV}}

\def\be{\begin{equation}}
\def\ee{\end{equation}}
\def\bea{\begin{eqnarray}}
\def\eea{\end{eqnarray}}

\begin{document}
\begin{flushright} DESY 99-002
\end{flushright}
\vspace{0.5in}
\title{Leptogenesis And Inflation\footnote{talk given at the International Workshop on Strong and Electroweak Matter (Copenhagen, 2-5 December 1998)}}

\author{D. Delepine}

\address{DESY,Notkestrasse 85, D-22607 Hamburg, Germany 
\\E-mail: delepine@mail.desy.de }


\maketitle\abstracts{In this talk, we studied the implication of the constraint on the reheating temperature coming from the gravitino problem on models of leptogenesis.  We point out that in supersymmetric extensions of the standard model, all existing models of neutrino masses and leptogenesis, except the one with right-handed singlet neutrinos are ruled out for a large range of the gravitino mass.
}
 Recent  announcement of experimental indications\cite{atm} for  very small neutrino masses implies an extension of the standard model (SM) where the right-handed neutrinos and  a mechanism to explain the smallness of the neutrino's masses have to be included.
In this talk, first, we review the different models of neutrino masses and  leptogenesis.
 Secondly, we evaluate the effects of the SM gauge interaction on leptogenesis.
 Finally, we discuss the gravitino constraint on the reheating temperature and its implications for the neutrino masses and leptogenesis models.
\section{Models of neutrino masses and leptogenesis}
All the models of neutrino masses
 require existence of new heavy particles $H$ with a mass $
M$ and lepton number violating interactions. At low energy this results in
an effective dimension-5 operator ${\cal O}=h^{2}{\frac{1}{M}}\ell _{L}\ell
_{L}\phi \phi $, where $\phi $ is the usual higgs doublet which gives masses
to the quarks and charged leptons.
 $H$ could be a $SU(2)_{L}$ singlet or a triplet
and it could be a fermion or a scalar, which gives four possible categories
of models for neutrino masses \cite{dim5}.
In the following, we briefly discuss the different classes of models:

 {\sl Right handed neutrinos:} The fermion content of the standard model is
extended to include one right handed neutrinos ($N_\alpha$, $\alpha = 1,2,3$%
) per generation, which are singlets under the SM  gauge group.
The right handed neutrinos have  Majorana masses and  Yukawa couplings 
with other leptons.
By the usual see-saw mechanism\cite{seesaw}, the left-handed
neutrinos get a small Majorana mass $M_\nu = - m_D^T {\frac{1 }{M_R}} m_D $.
The decay of $N_{\alpha }$ can generate the amount of lepton
asymmetry of the universe\cite{fy,lg}, if it satisfies the out-of-equilibrium condition,
$\Gamma _{N_{\alpha }}={\frac{h_{i\alpha }^{2}}{8\pi }}M_{\alpha }<H= 1.7%
\sqrt{g_{*}}{\frac{T^{2}}{M_{Pl}}}~~~~~{\rm at}~~T=M_{\alpha }$, where, $H $
is the Hubble constant, $g_{*}$ is the effective number of helicity states, $%
M_{Pl}$ is the Planck scale and $h_{i\alpha}$, the Yukawa coupling of the neutrinos.

{\sl Triplet higgs :} It is possible to extend the standard model to include
$SU(2)_L$ triplet higgs scalar fields\cite{triplet} ($\xi_a \equiv (1,3,-1)$, $a=1,2$ two
of them are required for $CP$ violation), whose relevant interactions are,
\begin{equation}
{\cal L} = M_a \xi_a^\dagger \xi_a + f_{ij} \xi_a^\dagger \ell_i \ell_j +
\mu \xi_a \phi \phi + h.c.
\end{equation}
The triplet higgs acquires a very tiny $%
vev$, which gives a Majorana mass to the neutrinos. 
 Lepton number violation comes
from the decays of the triplet higgs, $\xi_a$.
The one loop self-energy
diagrams interfere with the tree level decays to give $CP$ violation. If $%
M_2 > M_1$ and $\xi_1$ decays away from thermal equilibrium, {\it i.e.}, $%
\Gamma_1 
< H ~~{\rm at}%
~~T = M_1$, then a lepton asymmetry will be generated\cite{ma}.

{\sl Triplet Majorana fermions :} One can also extend the standard model to
include a $SU(2)_L$ triplet fermions, whose large Majorana masses breaks
lepton number. For all practical purposes they behave similar to those like
the right handed neutrinos and give neutrino masses through see-saw
mechanism\cite{tripletfermion}. Their decay can generate a lepton and hence
baryon asymmetry of the universe.

{\sl Radiative models :} It is possible to write down the effective
dimension-5 operator with $SU(2)_L$ singlet field ($\chi_a \equiv (1,1,-1)$,
$a=1,2$, two of these fields are required to get $CP$ violation) if there
are at least two higgs doublets ($\phi_a, a=1,2$). The neutrino masses
originate from radiative diagrams and hence are naturally small \cite{zee}.
 The relevant part of the Lagrangian for leptogenesis is given by 
\begin{equation}
{\cal L} = M^\chi_a \chi_a^\dagger \chi_a + f^\chi_{aij} \chi_a^\dagger
\ell_i \ell_j + \mu_\chi \chi_a \phi_1 \phi_2 + h.c.
\end{equation}
Among all the above four classes of models for neutrino masses, only models
with a singlet right-handed neutrino does not have any standard model gauge
interaction. In all the three other classes of models, the new particles
whose interactions break lepton number, transform non-trivially under the
standard model. We shall next study the consequences of the SM 
gauge interaction on leptogenesis. 

\section{SM gauge interaction and leptogenesis}
For simplicity, we shall consider a couple of generic heavy scalar $%
H_{a},a=1,2$, which couples to the standard model gauge bosons through gauge
interactions. In a supersymmetric model, the corresponding superpartner will
have similar gauge interactions with the gauginos and hence will suffer from
the same problem.
For the generation of a lepton asymmetry of the universe we
assume that the relevant part of the lagrangian is similar to that of
eqn(1). 
The generated  $CP$ asymmetry is given by $\eta $. We shall also assume that $%
M_{1}^{h}<M_{2}^{h}$, so that first $M_{2}^{h}$ decays and then the decay of
$M_{1}^{h}$ generates the lepton asymmetry of the universe.
The evolution of lepton number ($n_L = n_\ell - n_{\ell^c}$) is given by the
Boltzmann equation \cite{fry},
\begin{equation}
{\cal D} n_L = \eta \Gamma_H [n_H - n_H^{eq} ] - \left( \frac{n_L}{n_\gamma}
\right) n_H^{eq} \Gamma_H -2 n_\gamma n_L \langle \sigma_L |v| \rangle ,
\label{nl}
\end{equation}
where, the operator ${\cal D} \equiv \left[ \frac{{\rm d}}{{\rm d}t} + 3 H
\right]$; $n_H^{eq}$ is the equilibrium distribution of $H_1$ given by $%
n_{H}^{eq}=\frac{TM_{1}^{h2}}{2\pi^{2}}K_{2}(\frac{M_{1}^{h}}{T})$; $\Gamma_H
$ is the thermally-averaged decay rate of $H_a$; $n_\gamma$ is the photon
density and $\langle\sigma_L |v|\rangle$ is the thermally-averaged lepton
number violating scattering cross section.
The number density ($n_H$) of $H_1$ satisfies the Boltzmann equation,
\begin{equation}
{\cal D} n_H = - \Gamma_H (n_H - n_H^{eq}) + ({n_H}^2 - {n_H^{eq}}^2)
\langle \sigma_H |v| \rangle .  \label{nN}
\end{equation}
The second term on the right is the lepton number {\sl conserving}
thermally-averaged $H_1^\dagger + H_1 \to W_L + W_L$ scattering cross
section of the heavy particles $H_1$.
The  details of the computation can be found in ref(3). Here, we shall summarize the main results. 

For our analysis we shall thus assume, $\frac{\gamma \Gamma_{H(M_1^h=T)}}{H(M_1^h=T)}\ll 1$ at $T\sim M_{1}^{h}$ and thus $%
\eta <10^{-5}$. Taking the $SU(2)_{L}$ gauge coupling constant to be given
by the GUT coupling constant at the GUT scale, the effects of the SM gauge scattering  are presented in figure 1. 
\input{epsf.sty}
\begin{figure}[t]
\leavevmode
\par
\begin{center}
\mbox{\epsfxsize=7.cm\epsfysize=4.5cm\epsffile{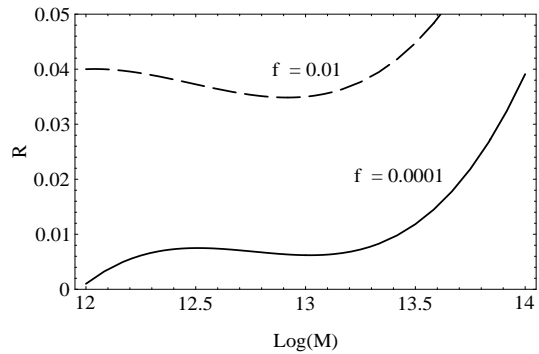}}
\end{center}
\caption{Lepton asymmetry of the universe for different masses of $H_{1}$,
when effects of gauge interaction is included. R is defined as $R= \frac{%
(n_{L}/s)_{with~~H_{1}^{\dagger }+H_{1}\to W_{L}+W_{L}}}{%
(n_{L}/s)_{without~~H_{1}^{\dagger }+H_{1}\to W_{L}+W_{L}}}$ and $%
f=f_{ij}^{h}$}
\label{fig1}
\end{figure}
So, for the allowed value of $\eta \leq 10^{-5}$
the lowest possible $H_{1}^{h}$ mass for the generation of enough lepton
asymmetry of the universe becomes
$M_{1}^{h}>O(10^{12})\gev.  \label{t12}$

\section{Gravitino problem and leptogenesis}

We have to keep in mind that leptogenesis can occur only after the end of
inflation.
In supersymmetric theories, the thermal production of massive
gravitinos restricts the beginning of the radiation-dominated era following
inflation except when the gravitino is very light\cite{pagels}. After the inflation a large number of gravitinos are
produced, which interact very weakly. The late decays of unstable gravitinos
can then modify the abundances of light elements causing inconsistency with
observation. In the other hand, stable gravitinos
may overclose the universe. This imposes a
upper bound on the reheating temperature $T_{RH}$~~\cite{linde,grav,ab,moroi}.

In the case of stable gravitinos, a limit on the $T_{RH}$ can be derived
from the closure limit of the universe\cite{plasma},
\begin{equation}
T_{RH}\leq 10^{10}\gev \times (\frac{m_{3/2}}{100 \gev}) \times
(\frac{1 \tev}{m_{\tilde{g}}(\mu)})^{2}
 \label{stable}
\end{equation}
with $m_{3/2}$ is the mass of the gravitino and $m_{\tilde{g}}(\mu)$ is
the running mass of the gluino. 
In the case of unstable gravitinos, the upper bound on the $T_{RH}$ depends
on the $m_{3/2}$. Essentially, one gets the followings constraints from
primordial nucleosynthesis\cite{moroi2}
\begin{equation}
T_{RH}\leq 10^{9}\gev~~~~~~~~~~~~~~~~~~~~~m_{3/2}< 1\tev
\end{equation}
\begin{equation} 
T_{RH}\leq 10^{12}\gev~~~~~~~1\tev < m_{3/2}<  5 \tev
\end{equation}

So, even for the stable
gravitino, when the bound on the reheating temperature is given by eqn(\ref
{stable}), it will not be possible to generate enough lepton asymmetry in
these scenarios, where the lepton number violating particle have got
standard model gauge interactions.

In summary, we point out that all the supersymmetric models of neutrino
masses, except for the one with singlet right handed neutrinos and
left-right symmetric models, may not be able to generate enough lepton
asymmetry of the universe consistently with the gravitino bound on the
reheating temperature in inflationary universe.
\section*{Acknowledgments}
This work was done in collaboration with Utpal Sarkar (Phys.Res.Lab., Ahmedabad).  We thank Prof. W. Buchm\"uller for useful comments and discussions.                             

\end{document}